# (Accepted Manuscript) TERIME: An Improved RIME Algorithm with Enhanced Exploration and Exploitation for Robust Parameter Extraction of Photovoltaic Models

Shi-Shun Chen, Yu-Tong Jiang, Wen-Bin Chen, Xiao-Yang Li[*]

School of Reliability and Systems Engineering, Beihang University, Beijing 100191, China

Email: css1107@buaa.edu.cn (Shi-Shun Chen), sy2314206@buaa.edu.cn (Yu-Tong Jiang), chenwenbin@buaa.edu.cn (Wen-Bin Chen), leexy@buaa.edu.cn (Xiao-Yang Li)

Corresponding author[*]: Xiao-Yang Li

## Abstract

Parameter extraction of photovoltaic (PV) models is crucial for the planning, optimization, and control of PV systems. Although some methods using meta-heuristic algorithms have been proposed to determine these parameters, the robustness of solutions obtained by these methods faces great challenges when the complexity of the PV model increases. The unstable results will affect the reliable operation and maintenance strategies of PV systems. In response to this challenge, an improved rime optimization algorithm with enhanced exploration and exploitation, termed TERIME, is proposed for robust and accurate parameter identification for various PV models. Specifically, the differential evolution mutation operator is integrated in the exploration phase to enhance the population diversity. Meanwhile, a new exploitation strategy incorporating randomization and neighborhood strategies simultaneously is developed to maintain the balance of exploitation width and depth. The TERIME algorithm is applied to estimate the optimal parameters of the single diode model, double diode model, and triple diode model combined with the Lambert-W function for three PV cell and module types including RTC France, Photo Watt-PWP 201 and S75. According to the statistical analysis in 100 runs, the proposed algorithm achieves more accurate and robust parameter estimations than other techniques to various PV models in varying environmental conditions. All of our source codes are publicly available at https://github.com/dirge1/TERIME.

| Abbreviations | |
|---|---|
| CLRao-1 | Comprehensive Learning Rao-1 |
| DDM | Double Diode Model |
| DE | Differential Evolution |
| DO | Dandelion Optimizer |
| ECM | Equivalent Circuit Model |
| IAE | Individual Absolute Error |
| MRIME | Modified RIME |
| NGO | Northern Goshawk Optimization |
| PV | Photovoltaic |
| RMSE | Root Mean Square Error |
| SD | Standard Deviation |
| SDM | Single Diode Model |
| SLCRIME | Sobol Local Cross RIME |
| SRIME | Strengthened RIME |
| TDM | Triple Diode Model |
| TERIME | Enhanced Exploration and Exploitation RIME |

# 1 Introduction

As the demand for energy grows worldwide, the shortage of fossil fuels and their environmental impact are becoming increasingly apparent. The effective development and utilization of renewable energy is a solution to avoid energy shortage and minimize environmental pollution [1]. Photovoltaic (PV) technology utilizing solar energy is the most promising renewable energy due to its high cost-effectiveness and excellent operational performance [2] and has been widely used in modern society [3-5]. Establishing reliable PV models is crucial for planning, optimizing, and controlling PV systems across various usage scenarios. However, developing physical models based on PV power generation mechanisms is challenging, since the relationship between the current and voltage of PV is implicit and nonlinear. In contrast, the Equivalent Circuit Model (ECM) simplifies the working mechanism of PV into electrical elements which are easier to analyze and understand [6]. Besides, ECMs are capable of adapting to various PV technologies and configurations, allowing for a wide range of applications in practical applications.

Typically, the primary ECMs employed for modeling the performance of PV systems are the Single Diode Model (SDM) [7], the Double Diode Model (DDM) [8], and the Triple Diode Model (TDM) [9]. The choice among these models is primarily determined by a balance between simplicity and accuracy. While the SDM is favored for its simplicity and ease of implementation, the DDM and TDM offer more detailed analysis, particularly valuable under low irradiance conditions [10]. Despite the effectiveness of ECMs in modeling PV systems, these models rely on parameters that are often unavailable in manufacturers' datasheets and vary significantly with environmental conditions [11]. Consequently, there has been a growing interest in accurate and robust parameter identification of PV models in varying environmental conditions.

In the literature, methods for estimating parameters in PV models can be generally divided into three categories: analytical methods [12], numerical methods [13], and meta-heuristic methods [14]. Analytical methods derive the analytical expressions for the unknown parameters by using three significant points from the manufacturers' datasheet: open circuit voltage, short circuit current, and maximum power point. However, recent findings indicate that the limited data available in the datasheet is inadequate to uniquely identify all the unknown parameters [15]. Therefore, researchers seek to extract parameters from the measured current-voltage curve (I-V curve) of the PV system to ensure the model accuracy [16]. Numerical methods, which employ the iterative method (e.g., Newton-Raphson approach) to extract parameters from the I-V curve, can theoretically determine PV parameters given sufficient data. Nevertheless, numerical methods often get stuck in local minima near the initial estimate, hindering the attainment of a global optimum [17]. Fortunately, meta-heuristic methods have shown excellent performance in extracting PV parameters, without the assumptions and initial estimates required by analytical and numerical methods. Therefore, many meta-heuristic algorithms have been utilized for PV parameter identification [18-20].

Although extensive research has been conducted on extracting PV parameters using meta-heuristic methods, accurate and reliable evaluation of these parameters remains challenging. As the complexity of the PV model increases, the robustness of meta-heuristic algorithms may degrade, greatly increasing computational costs [21]. Thus, the development of suitable meta-heuristic algorithms remains an open research question. In fact, the performance of meta-heuristic algorithms is highly dependent on the dual processes of exploration and exploitation [22]. Exploration is characterized by the investigation of completely new regions in a search space, whereas exploitation refers to visiting regions close to previously visited points [23]. RIME (rime optimization) is one of the latest meta-heuristic algorithms proposed by Su et al. [24] in 2023. It has shown robust exploration and exploitation capabilities compared to various basic meta-heuristic algorithms in multiple real-world problems. With its intuitive structure and no requirement for hyper-parameter tuning, it has garnered considerable attention and has already achieved good performance and robustness in various applications [25-27], including PV parameter extraction [28].

However, recent studies indicated that the RIME algorithm had flaws in the exploitation phase, causing it to become easily trapped in local optimums in high-dimensional optimization problems [29]. Besides, it also struggles to escape local optima during the original exploration phase, which significantly limits its effectiveness in practical applications [30]. To address the above problems, some researchers have sought to improve the RIME algorithm by enhancing either its exploration or exploitation phases [28, 31]. Nevertheless, these variants ignore the essential need to improve both exploration and exploitation capabilities simultaneously in the RIME algorithm. This may lead to the algorithm struggling with convergence or becoming prematurely trapped in local optima. In response to the above issue, Yuan et al. [29] proposed SLCRIME by incorporating the local optimal avoidance strategy and cross strategy. Specifically, the local optimal avoidance strategy boosted the exploratory ability based on two random agents and the cross strategy enhanced the interactive information exchange in the exploitation phase based on two other random agents. In Ref. [32], an improved version of RIME was developed featuring an interactive mechanism and a Gaussian diffusion strategy. The interactive mechanism employed two random agents and Levy flight mechanism to enhance the exploration, and the Gaussian diffusion strategy was introduced to boost the exploitation based on a random agent and the best agent.

Although existing studies have substantially enhanced the capability of the RIME algorithm, these variants still fall short in the exploitation phase, which seriously affects their robustness. To be specific, in the classic RIME algorithm [24], updates in the exploitation phase are based solely on the position of the current best agent. If this position is not the global optimum, all agents will be gradually assimilated and become trapped in local optima. This update strategy is preserved in Ref. [28], which limits their exploitation ability. In Refs. [29, 30], the authors sought to conduct the updates in the exploitation phase by exchanging information between random agents. While this strategy can be effective in escaping the local optima, it neglects the essence of the exploitation phase, i.e., the guidance of the best agent on other agents. As a result, the convergence speed will be reduced, computational expenses will be elevated, and the algorithm may fail to find the global optimum. Unlike the previous approach, the exploitation strategy was modified in Refs. [31, 32] by focusing on the neighborhood of the current best agent position. In fact, this strategy can be effective since the global optimum is sometimes located in the neighborhood of a local optimum, especially for the problem of PV parameter identification [33]. However, existing strategies still rely on random agents to determine the search range of the neighborhood, leading to an excessively large search area that hampers efficient and deep exploitation. Due to the flaws in the exploitation phase described above, these algorithms are struggling in estimating PV parameters reliably.

Motivated by the above challenges, an improved RIME algorithm with **E**nhanced **E**xploration and **E**xploitation (**T**riple **E**), termed TERIME, is proposed in this paper to enhance the robustness of PV parameter extraction for various PV models. In the TERIME, the randomization strategy and the neighborhood strategy are both incorporated into the exploitation phase. Additionally, inspired by the work of Ref. [28], a Differential Evolution (DE) mutation operator is integrated into the exploration phase to further enhance exploration capability. To show the effectiveness of the proposed approach, the parameter extraction results of three PV models (i.e., SDM, DDM, and TDM) using TERIME are compared with several state-of-the-art algorithms on three different datasets (RTC France, Photo Watt-PWP 201 and mono-crystalline S75). The main contributions of this paper can be summarized as follows:

- An improved RIME algorithm is developed by enhancing exploration and exploitation capabilities simultaneously for robust parameter identification of various PV models.
- The randomization and neighborhood strategies are both incorporated into the exploitation phase of the RIME algorithm for the first time to strike the balance of exploitation width and depth.
- The superior robustness of TERIME across various PV models and environmental conditions is demonstrated by comparisons with state-of-the-art meta-heuristic algorithms on three different PV systems.

The rest of this paper is organized as follows: In Section 2, the widely used PV models are introduced and the optimization problem is formulated. Then, the classic RIME algorithm and the proposed TERIME are presented in Sections 3 and 4, respectively. Next, Section 5 provides the experimental results. Finally, Section 6 concludes the paper.

## 2  PV Models and Optimization Problem Formulation

### 2.1  Single Diode Model

The equivalent circuit of the SDM is illustrated in Fig. 1 (a). According to the Kirchhoff's law, the output current of the SDM can be calculated as:

$$I = I_{ph} - I_d - I_{sh}, \tag{1}$$

where $I_{ph}$ is the generated photoelectric current; $I_d$ represents the current flowing through the diode; and $I_{sh}$ refers to the current flowing through the parallel resistance $R_{sh}$. By applying the Shockley diode equation, $I_d$ can be derived as:

$$I_d = I_o \left[ \exp\left(\frac{V + IR_s}{nV_t}\right) - 1 \right], \tag{2}$$

$$V_t = \frac{kT}{q}, \tag{3}$$

where $I_o$ denotes the reverse saturation current in the diode; $R_s$ is the series resistance; $V$ is the output voltage; $V_t$ is the thermal voltage represented as Eq. (3); $k$ is the Boltzmann constant ($1.380649 \times 10^{-23}$ J·K$^{-1}$); $q$ is the electron charge ($1.602176634 \times 10^{-19}$ C); and $T$ is the temperature of the PV cell in kelvin.

$I_{sh}$ in Eq. (1) can be computed as:

$$I_{sh} = \frac{V + IR_s}{R_{sh}}. \tag{4}$$

Then, based on Eqs. (1)-(4), the output current $I$ can be described as:

$$I = I_{ph} - I_o \left[ \exp\left(\frac{V + IR_s}{nV_t}\right) - 1 \right] - \frac{V + IR_s}{R_{sh}}. \tag{5}$$

In order to describe the performance of a PV cell by the SDM, there are five unknown parameters ($I_{ph}$, $I_o$, $n$, $R_s$ and $R_{sh}$) to be determined.

### 2.2  Double Diode Model

Fig. 1 (b) shows the equivalent circuit of the DDM. Compared to the SDM, the DDM takes into account the influence of charge carrier recombination loss on the depletion region [34]. Similar to the derivation of the SDM, the output current of the DDM can be formulated as:

$$I = I_{ph} - I_{o1} \left[ \exp\left(\frac{V + IR_s}{n_1 V_t}\right) - 1 \right] - I_{o2} \left[ \exp\left(\frac{V + IR_s}{n_2 V_t}\right) - 1 \right] - \frac{V + IR_s}{R_{sh}}, \tag{6}$$

where $I_{o1}$ and $I_{o2}$ are the reverse saturation current of the two diodes; and $n_1$ and $n_2$ denote the ideality factor of the two diodes. In the DDM, there are seven unknown parameters ($I_{ph}$, $I_{o1}$, $I_{o2}$, $n_1$, $n_2$, $R_s$ and $R_{sh}$) to be identified, implying higher dimensionality and more computation time with respect to the SDM.

### 2.3  Triple Diode Model

The equivalent circuit of the TDM is presented in Fig. 1 (c). In comparison to the DDM, the TDM can further consider the recombination loss in defect regions and grain sites [35]. Its output current can be written as:

$$I = I_{ph} - I_{o1}\left[\exp\left(\frac{V+IR_s}{n_1 V_t}\right)-1\right] - I_{o2}\left[\exp\left(\frac{V+IR_s}{n_2 V_t}\right)-1\right] - I_{o3}\left[\exp\left(\frac{V+IR_s}{n_3 V_t}\right)-1\right] - \frac{V+IR_s}{R_{sh}}, \quad (7)$$

where $I_{o3}$ and $n_3$ are the reverse saturation current and the ideality factor of the third diode, respectively. The TDM is the most complicated model with nine unknown parameters ($I_{ph}$, $I_{o1}$, $I_{o2}$, $I_{o3}$, $n_1$, $n_2$, $n_3$, $R_s$ and $R_{sh}$), requiring the highest computational cost.

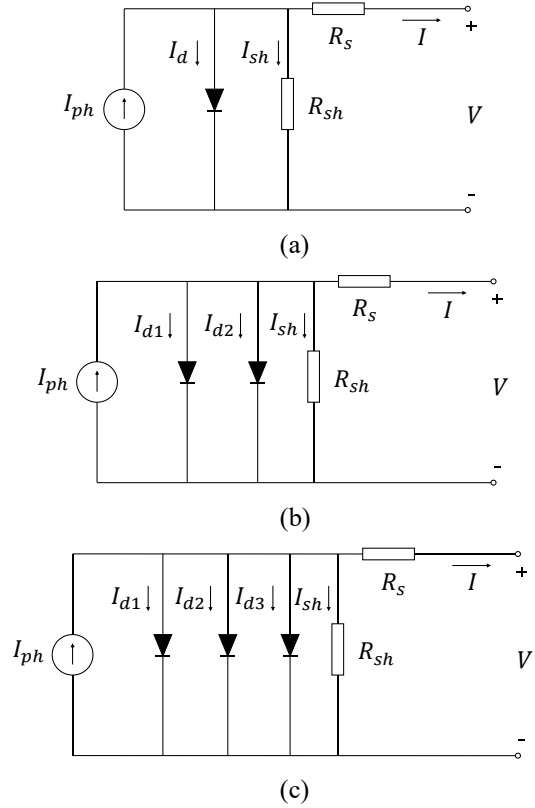

Fig. 1 The equivalent circuit of the PV model: (a) the SDM; (b) the DDM; (c) the TDM.

2.4 Optimization Problem Formulation

The purpose of the PV model parameter extraction is to make the constructed I-V curve based on the selected PV model as consistent as possible with the measured one. In general, the most commonly used and effective objective function is to minimize the Root Mean Square Error (RMSE) [36], which can be expressed as:

$$RMSE = \sqrt{\frac{1}{N_m}\sum_{i=1}^{N_m}\left[I_{cal,i}\left(V_{meaure,i},\boldsymbol{\Theta}\right) - I_{meaure,i}\right]^2}, \quad (8)$$

where $N_m$ represents the number of the measured points in the I-V curve; $I_{meaure,i}$ and $V_{meaure,i}$ are the measured output current and voltage of the $i^{th}$ measured point, respectively; $I_{cal,i}$ denotes the calculated output current of the $i^{th}$ measured point by the PV model given $V_{meaure,i}$; and $\boldsymbol{\Theta}$ are the unknown parameters that need to be estimated.

Since the PV models (5)-(7) are nonlinear implicit transcendental equations, it is difficult to solve them directly. In this paper, the Lambert W function is used to obtain the calculated current due to its superior accuracy and efficiency [37]. The Lambert W function, denoted as $W(x)$, is a multi-branched function defined as the set of functions satisfying the Eq. (9) for any complex number $x$.

$$W(x)e^{W(x)} = x. \tag{9}$$

Then, for the SDM, Eq. (5) can be rewritten as:

$$I = \frac{R_{sh}(I_{ph} + I_o) - V}{R_{sh} + R_s} - \frac{V_t}{R_s} nW(\beta), \tag{10}$$

where

$$\beta = \frac{I_o R_s R_{sh}}{nV_t(R_s + R_{sh})} \exp\left(\frac{R_{sh}(R_s I_{ph} + R_s I_o + V)}{nV_t(R_s + R_{sh})}\right). \tag{11}$$

For the DDM:

$$I = \frac{R_{sh}(I_{ph} + I_{o1} + I_{o2}) - V}{R_{sh} + R_s} - \frac{V_t}{R_s}\left[n_1 W(\beta_1) + n_2 W(\beta_2)\right], \tag{12}$$

where

$$\beta_1 = \frac{I_{o1} R_s R_{sh}}{n_1 V_t(R_s + R_{sh})} \exp\left(\frac{R_{sh}(R_s I_{ph} + R_s I_{o1} + V)}{n_1 V_t(R_s + R_{sh})}\right), \tag{13}$$

$$\beta_2 = \frac{I_{o2} R_s R_{sh}}{n_2 V_t(R_s + R_{sh})} \exp\left(\frac{R_{sh}(R_s I_{ph} + R_s I_{o2} + V)}{n_2 V_t(R_s + R_{sh})}\right). \tag{14}$$

For the TDM:

$$I = \frac{R_{sh}(I_{ph} + I_{o1} + I_{o2} + I_{o3}) - V}{R_{sh} + R_s} - \frac{V_t}{R_s}\left[n_1 W(\beta_1) + n_2 W(\beta_2) + n_3 W(\beta_3)\right], \tag{15}$$

where

$$\beta_3 = \frac{I_{o3} R_s R_{sh}}{n_3 V_t(R_s + R_{sh})} \exp\left(\frac{R_{sh}(R_s I_{ph} + R_s I_{o3} + V)}{n_3 V_t(R_s + R_{sh})}\right), \tag{16}$$

It should be noted that a PV module consisting of several cells connected in series can also be expressed by Eqs. (5)-(7). The only difference is the transformation of Eq. (3) into:

$$V_t = \frac{N_s kT}{q}, \tag{17}$$

where $N_s$ is the number of cells connected in series.

## 3 RIME Algorithm

The RIME algorithm is a physics-based meta-heuristic optimization technique inspired by the natural process of rime formation [24]. It distinguishes the growth patterns of soft-rime and hard-rime under different wind speed. The optimization procedure is shown below.

### 3.1 Rime Population Initialization

Similar to other population-based optimization techniques, the RIME algorithm starts by generating the initial population $X$. Specifically, the rime population consists of $N$ rime agents, and each agent is randomly positioned within the search space to form the initial population, which can be mathematically expressed as:

$$X = \begin{bmatrix} x_{11} & x_{12} & \cdots & x_{1D} \\ x_{21} & x_{22} & \cdots & x_{2D} \\ \vdots & \vdots & \ddots & \vdots \\ x_{N1} & x_{N2} & \cdots & x_{ND} \end{bmatrix}, \tag{18}$$

$$x_{ij} = LB_j + r_0 \times (UB_j - LB_j), i \in \{1,2,\ldots,N\}, j \in \{1,2,\ldots,D\}, \tag{19}$$

where $D$ represents the dimension of the optimized problem; $i$ and $j$ are the ordinal numbers that denote the agents and the particles, respectively; $r_0$ is a value randomly selected ranging between 0 and 1; and $UB_j$ and $LB_j$ represent the upper and lower boundaries of the $j^{th}$ particle, respectively.

3.2  Soft-Rime Search Strategy

Under breezy conditions, the development of soft-rime is entirely random and slow. Based on this phenomenon, the RIME algorithm introduces a soft-rime search strategy. This strategy can efficiently span the entire search space and avoid becoming trapped in local optima. The location of each particle can be formulated as:

$$x_{new,i}^j = x_{best}^j + r_1 \cdot \cos\theta \cdot \beta \cdot \left(h \cdot (UB_i^j - LB_i^j) + LB_i^j\right), r_2 \leq E, \tag{20}$$

where $x_{new,i}^j$ represents the new position update for the $j^{th}$ particle of the $i^{th}$ rime agent; $x_{best}^j$ indicates the position of the $j^{th}$ particle of the best-performing rime agent currently; $r_1$ is a control parameter that influences the direction of particle movement, which is randomly selected from -1 to 1; $r_2$ is another random number ranging from 0 to 1; $\theta$ adjusts according to the number of iterations, which can be calculated by Eq. (21); $\beta$ is a variable determined by Eq. (22), illustrating the effect of environmental conditions on the process; $E$ denotes a factor affecting the probability of condensation as depicted in Eq. (23).

$$\theta = \pi\left(\frac{t}{10 \cdot T_{max}}\right), \tag{21}$$

$$\beta = 1 - \left[\frac{w \cdot t}{T_{max}}\right]/w, \tag{22}$$

$$E = \sqrt{\frac{t}{T_{max}}}, \tag{23}$$

where $T_{max}$ denotes the maximum number of iterations; $t$ indicates the present iteration number; $[\cdot]$ represents the rounding operator; $w$ is assigned as 5.

3.3  Hard-Rime Puncture Mechanism

Hard-rime forms under strong gale conditions. Its growth pattern is simpler and more regular compared to that of soft-rime. The RIME algorithm leverages this phenomenon and introduces the hard-rime puncture mechanism, which can effectively enhance the convergence and avoid local optima. This mechanism can be mathematically expressed as:

$$x_{new,i}^j = x_{best}^j, r_3 < F_{norm}(x_i), \tag{24}$$

where $F_{norm}(x_i)$ represents the normalized fitness value of the $i^{th}$ search agent with respect to all agents, which determines the selection probability of the specific agent; $r_3$ is a random number ranging between 0 and 1.

3.4  Positive Greedy Selection Mechanism

Through the positive greedy selection mechanism, the fitness value of the updated search agent is evaluated in comparison to the previous agent. When the fitness of the updated agent exceeds that of the previous agent, it replaces the previous agent, updating both the agent and its fitness value. This approach incrementally improves the quality of the search agents, ensuring continuous population improvement with each iteration.

## 4 Proposed TERIME Algorithm

### 4.1 Enhanced Exploration Approach

As mentioned in the introduction, the exploration phase of the RIME algorithm formulated in Eq. (20) is associated with the current best agent. If the algorithm becomes trapped in a local optimum, it will be hard to escape. To alleviate this problem, inspired by the work of Ref. [28], we introduce DE mutation operators. DE mutation operators are common strategies for enhancing the population diversity of meta-heuristic algorithms [38]. By introducing variations through perturbation mechanism, these operators enable the agents to explore the search space more thoroughly and avoid premature convergence. Among these operators, the DE/rand/1 operator is employed in this paper due to its simplicity, effectiveness, and high randomness [39], which can be expressed as:

$$x_{new,i} = x_i + \varphi \cdot (x_a - x_b) \tag{25}$$

where $x_a$ and $x_b$ are two randomly selected agents from the population; and $\varphi$ is the mutation factor randomly generated between 0 and 1. Then, we rewrite the original exploration phase Eq. (20) as follows:

$$\begin{cases} x_{new,i} = x_i + \varphi \cdot (x_a - x_b), r_4 > 0.5 \\ x_{new,i}^j = x_{best}^j + r_1 \cdot \cos\theta \cdot \beta \cdot \left( h \cdot (UB_i^j - LB_i^j) + LB_i^j \right), r_2 \le E \text{ and } r_4 \le 0.5 \\ x_{new,i} = x_i, r_2 > E \text{ and } r_4 \le 0.5 \end{cases} \tag{26}$$

where $r_4$ is a random number ranging between 0 and 1. By combing the DE/rand/1 operator, updates of some agents in the exploration phase will not rely on the current optimal agent, thus enabling the algorithm to escape local optima.

### 4.2 Enhanced Exploitation Strategy

In the exploitation phase of the RIME algorithm, updates are based solely on the position of the current best agent as described in Eq. (24). If this position is not the global optimum, all agents will eventually be assimilated, leading to entrapment in a local optimum. To address this issue, the crossover and Gaussian exploitation strategies are developed and integrated.

The crossover strategy is a randomization exploitation strategy, which can be described as [29]:

$$x_{new,i}^j = x_{i,j}^j + C \cdot (x_c^j - x_d^j) \tag{27}$$

$$C = \left( \cos\left(\frac{\pi t}{T_{max}}\right) + 1 \right) \cdot \left( 1 - \frac{t}{2T_{max}} \right) \tag{28}$$

where $C$ is the crossover factor derived from (28); $x_c^j$ and $x_d^j$ are the $j^{th}$ particle of two randomly selected agents from the population. The key difference between Eq. (25) and Eq. (27) is the control factor: $\varphi$ in Eq. (25) is a random number, while $C$ in Eq. (27) decreases as the iteration number increases.

On the other hand, the Gaussian exploitation strategy is a neighborhood exploitation strategy, which can be derived as:

$$x^j_{new,i} = G\left(x^j_{best}, \sigma x^j_{best}\right) \tag{29}$$

where $G\left(x^j_{best}, \sigma x^j_{best}\right)$ denotes a normally distributed random number with a mean of $x^j_{best}$ and a standard deviation of $\sigma x^j_{best}$; and $\sigma$ is the variation coefficient, and its value is determined by the specific problem. In this paper, $\sigma$ in Eq. (29) is set as 0.001. Then, the original exploitation phase Eq. (24) can be rewritten as:

$$\begin{cases} x^j_{new,i} = x^j_{i,j} + C \cdot \left(x^j_c - x^j_d\right), r_3 < F_{norm}(x_i) \text{ and } r_5 \geq 0.5 \\ x^j_{new,i} = G\left(x^j_{best}, \sigma x^j_{best}\right), r_3 < F_{norm}(x_i) \text{ and } r_5 < 0.5 \\ x^j_{new,i} = x^j_i, r_3 \geq F_{norm}(x_i) \end{cases} \tag{30}$$

where $r_5$ is a random number ranging between 0 and 1. Compared to other variants of RIME algorithm [28-32], the combination of crossover and Gaussian exploitation strategies facilitates information exchange among agents and sufficiently exploits the neighborhood of the current best agent, which can assist the algorithm in avoiding local optima and enhances the convergence speed.

4.3 Application Procedure of TERIME

With the consideration of the above improvements, the pseudocode of TERIME is presented in Algorithm 1, and the flowchart of TERIME is illustrated in Fig. 2. It is worth mentioning that after the exploration and exploitation phases, the new position of the agents might fall outside the boundary. In order to alleviate the early convergence issue resulting from the clustering of agents near the boundaries of the search area, we adopt the strategy advised in Ref. [40] to adjust their positions, which can be expressed as:

$$x^j_{new} = \frac{UB_j + LB_j}{2} + \frac{UB_j - LB_j}{2} \cdot (2\gamma - 1) \tag{31}$$

where $\gamma$ is a random number between 0 and 1.

**Algorithm 1**: Pseudocode of TERIME.
**Input**:
1. $T_{max}$: Maximum number of iterations
2. $N$: Number of agents in the population
3. $LB$, $UB$: Lower and upper bounds of variables
4. $f(\ )$: Objective function

**Output**:
1. The optimal solution and its best fitness value

**Procedure**:
Initialization
1. Generate the initial population of TERIME using Eq. (18)
2. Evaluate the fitness of the population using Eq. (8)
3. Calculate $F_{norm}$ of each agent using min-max normalization
4. **While** $t<T_{max}$ **do**
Exploration
5.     Update $\theta$, $\beta$, $E$, $C$ using Eqs. (21), (22), (23) and (28)
6.     **For** $i=1$ to $N$ **do**
7.         Generate random numbers $r_2$ and $r_4$
8.         **If** $r_4<0.5$ **do**
9.             Update the position of agent $i$ using Eq. (25)
10.        **Else do**

| | |
|---|---|
| 11. | **If** $r_2 < E$ **do** |
| 12. | Update the position of agent $i$ using Eq. (20) |
| 13. | **End if** |
| 14. | **End if** |
| 15. | **End for** |
| Exploitation | |
| 16. | **For** $i=1$ to $N$ **do** |
| 17. | Generate random numbers $r_3$ and $r_5$ |
| 18. | **If** $r_3 < F_{norm}(x_i)$ **do** |
| 19. | **If** $r_5 < 0.5$ **do** |
| 20. | Update the position of agent $i$ using Eq. (29) |
| 21. | **Else do** |
| 22. | Update the position of agent $i$ using Eq. (27) |
| 23. | **End if** |
| 24. | **End if** |
| 25. | **End for** |
| Boundary check and update | |
| 26. | Perform boundary check and adjustment using Eq. (31) |
| 27. | Evaluate the fitness of the updated population using Eq. (8) |
| 28. | Replace any agent with the updated one if the fitness is improved |
| 29. | **End while** |

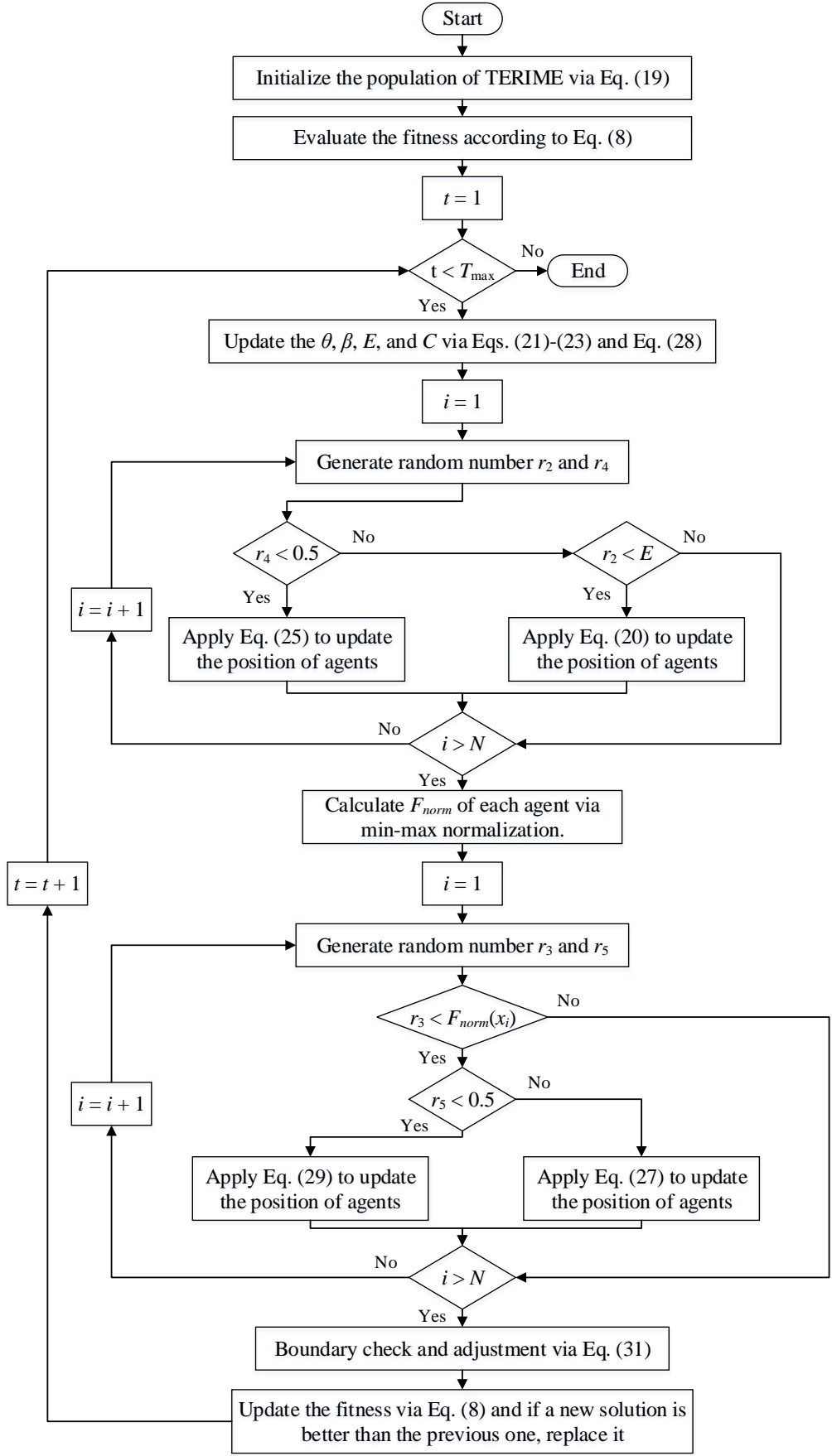

Fig. 2 Flow chart of TERIME.

## 5 Experimental Results

### 5.1 Dataset Description and Validation Schemes

In this section, I-V characteristics of three PV systems are introduced to implement the proposed TERIME algorithm. These PV datasets are widely used in the literature to assess parameter estimation techniques for PV models [40]. Brief introductions to these PV datasets is given as follows.

*Dataset 1*: a commercial solar cell RTC France, 57 mm in diameter, with experimental I-V data recorded at 33°C under full illumination.

*Dataset 2*: a poly-crystalline PV module called Photo Watt-PWP 201, which consists of 36 series-connected cells. Its experimental I-V characteristic was measured at 45°C and 1000 W/m² irradiance.

*Dataset 3*: a poly-crystalline S75 PV module, composed of 36 cells in series, and its experimental I-V data were measured under varying environmental conditions. Firstly, the module was tested at a standard temperature of 25°C under five irradiance levels: 1000 W/m², 800 W/m², 600 W/m², 400 W/m², and 200 W/m². Then, the test was performed at three different temperatures (20°C, 40°C, and 60°C) under a standard irradiance of 1000 W/m².

The variation range of the parameters in SDM, DDM, and TDM of these three PV systems is presented in Table 1 [40], where $I_{sc}$ is the short circuit current of the S75 PV module and its calculation is described in the Appendix.

Table 1 Lower and upper limits for the parameters of the three PV systems.

| Parameters | RTC France | | PWP 201 | | S75 | |
|---|---|---|---|---|---|---|
| | LB | UB | LB | UB | LB | UB |
| $I_{ph}$ / A | 0 | 1 | 0 | 2 | 0 | $2I_{sc}$ |
| $I_o, I_{o1}, I_{o2}, I_{o3}$ / $\mu$A | 0 | 1 | 0 | 10 | 0 | 1 |
| $R_s$ / $\Omega$ | 0 | 0.5 | 0 | 2 | 0 | 2 |
| $R_{sh}$ / $\Omega$ | 0 | 100 | 0 | 2000 | 0 | 5000 |
| $n, n_1, n_2, n_3$ | 1 | 2 | 1 | 2 | 1 | 4 |

To verify the effectiveness of TERIME, two rules are followed to choose the representative competing algorithms:

- Variants of the RIME algorithm are considered, which include MRIME [28], SLCRIME [29] and SRIME [31]. These variants strengthen exploration [28], exploitation [29], or both [31], helping to demonstrate the advantages of the proposed enhanced strategy.
- State-of-the-art algorithms for PV parameter estimation are considered, which include DO [41], NGO [42], and CLRao-1 [43]. These techniques, proposed in the last three years, are noted for their robust performance and are well-suited as benchmarks for comparison.

In addition, the I-V characteristics of the PV models based on the parameters extracted by TERIME are compared with the measured values to further demonstrate its effectiveness.

In practical applications, computational cost is the primary focus in PV parameter estimation. For fair comparison, all the algorithms are configured with identical computational limits, setting the maximum objective function evaluations $E_{max}$ at 100000 and the population size $N$ at 20. In TERIME, $E_{max}=T_{max}$ since the objective function evaluation is performed once per iteration. Besides, since results from these algorithms vary with each run, all the algorithms are independently run 100 times on each PV model. The Maximum (Max), Mean, Minimum (Min), and Standard Deviation (SD) RMSE values of 100 runs are then calculated to assess the performance and robustness of these algorithms.

5.2 Results of RTC France PV Cell

According to the settings in Section 5.1, the RMSE values of SDM, DDM and TDM parameter extraction for RTC France in 100 runs are shown in Table 2. The PV parameters corresponding to the smallest RMSE for different algorithms are given in the supplementary material. From Table 2, the following findings could be given:

- For the SDM, a robust global optimum 7.730063e-4 can be obtained by TERIME, MRIME and CLRao-1 in all the 100 runs.
- For the DDM, TERIME delivers the best results for Mean, Max and SD values, significantly outperforming other algorithms. MRIME and CLRao-1 can give a global optimum in 100 runs, but their robustness is inferior.
- For the TDM, MRIME gives a best Min value 5.843708e-4 in 100 runs, followed by TERIME. However, regarding Mean and Max values, TERIME achieves the best results and is markedly superior to those of other techniques. Although SLCRIME and NGO give better SD than that of TERIME, their Min, Mean and Max values are inferior to those of TERIME.

Table 2 Comparison of RMSE results from 100 runs for the SDM, DDM and TDM parameter extraction of RTC France.

| Algorithms | Model | Min / $10^{-4}$ | Mean / $10^{-4}$ | Max / $10^{-4}$ | SD |
|---|---|---|---|---|---|
| TERIME | | **7.730063** | **7.730063** | **7.730063** | **1.0e-17** |
| RIME | | 7.731401 | 15.888173 | 20.833180 | 0.00050 |
| SLCRIME | | **7.730063** | **7.730063** | 7.730084 | 2.5e-10 |
| SRIME | SDM | 7.743956 | 14.817024 | 23.041004 | 0.00041 |
| MRIME | | **7.730063** | **7.730063** | **7.730063** | 2.8e-17 |
| DO | | 7.730093 | 9.137755 | 20.574846 | 0.00017 |
| NGO | | 7.732488 | 7.777358 | 7.989294 | 4.5e-06 |
| CLRao-1 | | **7.730063** | **7.730063** | **7.730063** | 2.0e-17 |
| TERIME | | **6.745134** | **6.745134** | **6.745134** | **2.0e-17** |
| RIME | | 8.011388 | 29.400522 | 44.194571 | 0.00113 |
| SLCRIME | | 6.794156 | 7.585225 | 8.032144 | 2.0e-05 |
| SRIME | DDM | 7.679083 | 13.633113 | 23.849559 | 0.00038 |
| MRIME | | **6.745134** | 6.995867 | 7.952999 | 3.8e-05 |
| DO | | 6.787838 | 8.894699 | 28.007019 | 0.00029 |
| NGO | | 7.322259 | 7.974473 | 9.642399 | 2.8e-05 |
| CLRao-1 | | 6.745135 | 7.343943 | 7.730063 | 2.6e-05 |
| TERIME | TDM | 5.932588 | **6.455588** | **7.298956** | 6.3e-05 |

| | | | | |
|---|---|---|---|---|
| RIME | 7.144043 | 21.347482 | 55.690168 | 0.00136 |
| SLCRIME | 6.184865 | 7.508216 | 7.954176 | **2.6e-05** |
| SRIME | 7.221689 | 12.212359 | 23.507644 | 0.00032 |
| MRIME | **5.843708** | 6.625793 | 20.831811 | 0.00021 |
| DO | 6.526195 | 9.382817 | 29.922138 | 0.00035 |
| NGO | 7.320781 | 8.123815 | 10.323165 | 4.6e-05 |
| CLRao-1 | 5.944958 | 7.309326 | 20.832507 | 0.00014 |

Then, Fig. 3 illustrates the RMSE box plots for the three excellent algorithms, i.e., TERIME, MRIME, and CLRao-1. It is evident that TERIME exhibits robust results with good performance for all the PV models, especially for the DDM. However, MRIME and CLRao-1 show reduced performance when the complexity of the PV model increases.

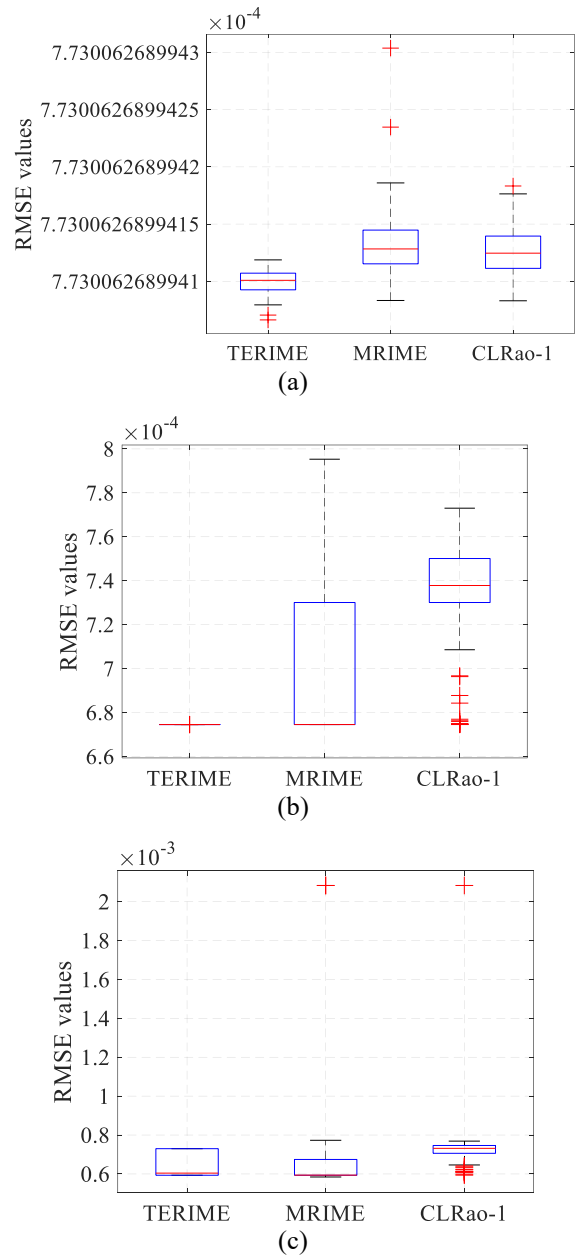

Fig. 3 RMSE box plots for three excellent algorithms for parameter extraction of RTC France: (a) SDM;

(b) DDM; (c) TDM.

Next, Fig. 4 illustrates the average convergence of all the algorithms for the SDM, DDM and TDM parameter extraction. From Fig. 4, it can be found that:

- For the SDM, TERIME converges fastest to the neighborhood of the global optimum in the first 20000 iterations.
- For the DDM, MRIME has the fastest convergence speed in the first 20000 iterations, while TERIME outperforms all the algorithms after 40000 iterations.
- For the TDM, MRIME has the fastest convergence speed in the first 20000 iterations, followed by CLRao-1 and TERIME. However, MRIME is surpassed by TERIME after 70000 iterations.
- Among all the algorithms, RIME and SRIME are easily trapped in the local optimum. NGO, DO and SLCRIME converges slowly and fails to reach the global optima for the DDM and TDM. Compared to the above algorithms, TERIME, MRIME and CLRao-1 have better performance.

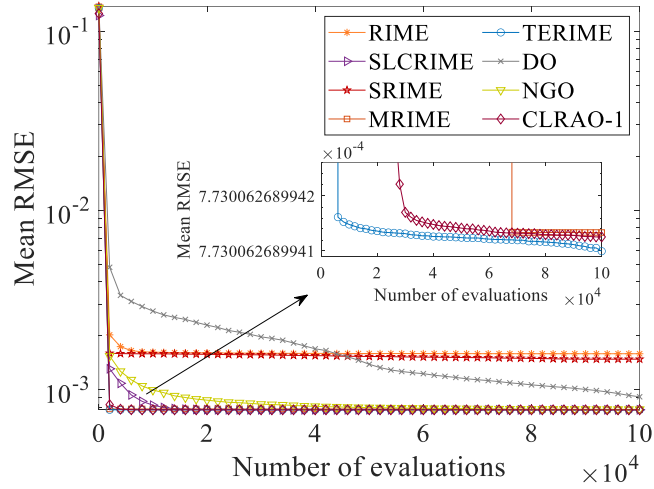

(a)

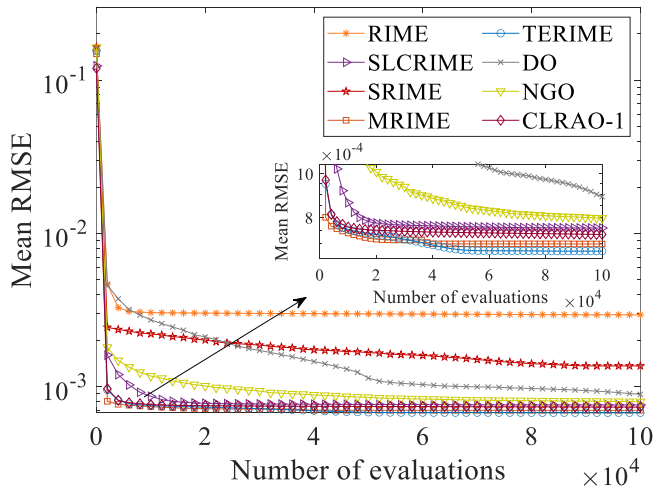

(b)

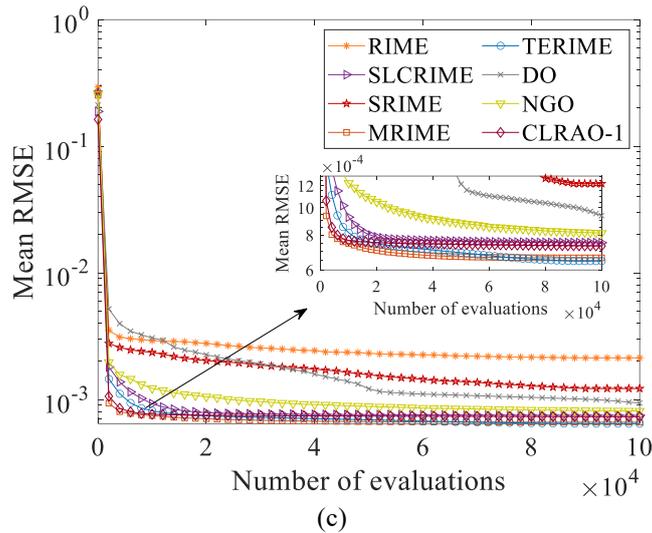

(c)

Fig. 4 Average convergence performance from 100 runs for parameter extraction of RTC France: (a) SDM; (b) DDM; (c) TDM.

In summary, the ranking of all the algorithms for DDM and TDM parameter extraction of RTC France is illustrated in Fig. 5. A smaller ranking indicates better performance. As shown in Fig. 5, we can conclude that TERIME is a competitive algorithm with satisfactory performance and excellent robustness. Although TERIME's Min ranking is worse than MRIME in TDM parameter extraction, TERIME achieves the best Mean and Max ranking in both DDM and TDM parameter extraction.

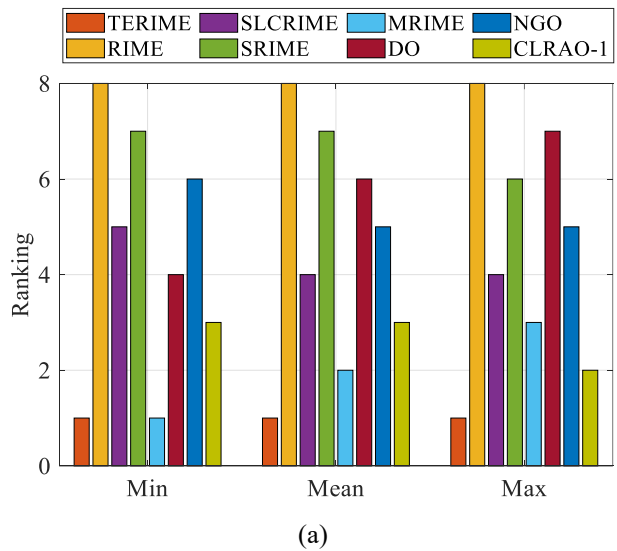

(a)

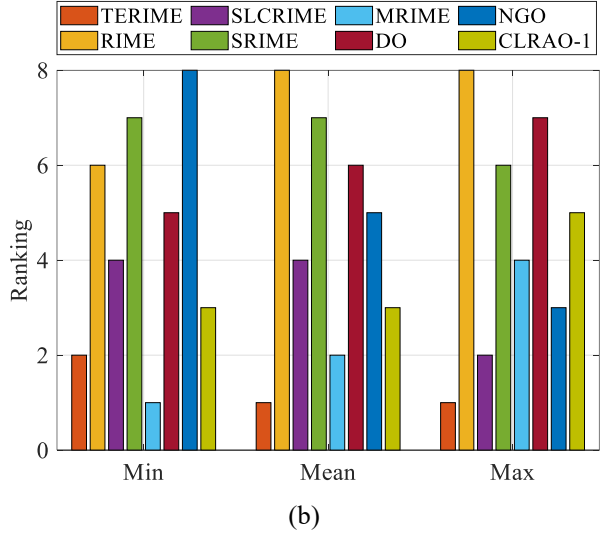

(b)

Fig. 5 The ranking of all the algorithms for parameter extraction on RTC France: (a) DDM; (b) TDM.

Subsequently, we focus on the goodness of fit between the calculated data obtained by the PV model and the measured one, since an accurate I-V characteristic is what we focus on. In order to quantify the error margins between the measured and calculated data, the Individual Absolute Error (IAE) for the output current is computed using Eq. (32). Then, the IAEs of SDM, DDM and TDM for the RTC France obtained by the TERIME are illustrated in Fig. 6.

$$\text{IAE}_i = \left| I_{cal,i}\left(V_{meaure,i}, \theta\right) - I_{meaure,i} \right|. \tag{32}$$

As can be seen in Fig. 6, the IAE for all the measured data remains below 1.6e-3 regardless of the PV model type, and is below 1e-3 for most data, indicating the effectiveness of TERIME. Moreover, TDM has the smallest IAE overall, followed by DDM and finally SDM, demonstrating that increasing the number of diodes enhances modeling accuracy in this case.

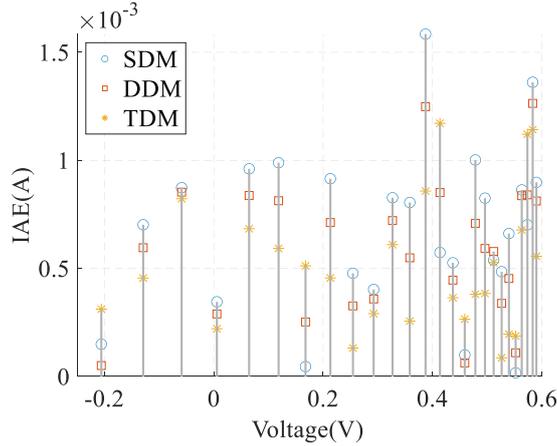

Fig. 6 IAEs of SDM, DDM and TDM for the RTC France obtained by the TERIME.

Furthermore, the calculated I-V curves from the SDM, DDM, and TDM using TERIME are compared with the measured one, as shown in Fig. 7. It can be deduced that the optimal parameters estimated by TERIME are very close to the actual cell parameters, since the calculated I-V curves fit nearly all the measurements.

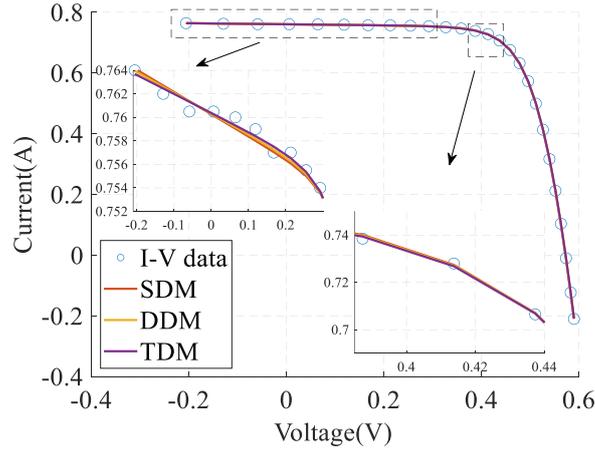

Fig. 7 Comparison of measured I-V curve and calculated ones by the SDM, DDM and TDM using TERIME for the RTC France.

5.3 Results of PWP 201 PV Module

The RMSE values of the SDM, DDM and TDM parameter extraction for the PWP 201 using different algorithms in 100 runs are shown in Table 3. The PV parameters corresponding to the smallest RMSE for different algorithms are given in the supplementary material. It can be seen that a robust globally optimal value of 1.980210e-3, 1.235854e-3 and 1.235854e-3 can be presented by TERIME in all the 100 runs for all the PV models. Although the performance of MRIME and CLRao-1 is robust for the SDM, their robustness declines as PV model complexity increases.

Table 3 Comparison of RMSE results from 100 runs for SDM, DDM and TDM parameter extraction of PWP 201.

| Algorithms | Model | Min / $10^{-3}$ | Mean / $10^{-3}$ | Max / $10^{-3}$ | SD |
|---|---|---|---|---|---|
| TERIME | | **1.980210** | **1.980210** | **1.980210** | **1.3e-17** |
| RIME | | 1.982655 | 2.881791 | 3.779420 | 0.00064 |
| SLCRIME | | **1.980210** | 1.980211 | 1.980215 | 1.0e-09 |
| SRIME | SDM | 1.983797 | 2.632392 | 3.727036 | 0.00050 |
| MRIME | | **1.980210** | **1.980210** | **1.980210** | 2.9e-17 |
| DO | | 1.980221 | 2.080552 | 2.331327 | 8.6e-05 |
| NGO | | 1.980507 | 1.985768 | 2.051930 | 7.9e-06 |
| CLRao-1 | | **1.980210** | **1.980210** | **1.980210** | 4.8e-17 |
| TERIME | | **1.235854** | **1.235854** | **1.235854** | **1.3e-17** |
| RIME | | 1.553085 | 3.806825 | 6.910481 | 0.00158 |
| SLCRIME | | 1.325982 | 1.507803 | 1.616501 | 6.1e-05 |
| SRIME | DDM | 1.788223 | 2.525695 | 3.574089 | 0.00039 |
| MRIME | | **1.235854** | 1.257171 | 1.977165 | 0.00010 |
| DO | | 1.299107 | 1.992142 | 2.445353 | 0.00018 |
| NGO | | 1.393959 | 1.949974 | 2.369973 | 0.00012 |
| CLRao-1 | | 1.243882 | 1.427011 | 1.980210 | 0.00016 |
| TERIME | | **1.235854** | **1.236097** | **1.259446** | **2.3e-06** |
| RIME | | 1.475659 | 3.093056 | 8.945386 | 0.00141 |
| SLCRIME | | 1.319558 | 1.539146 | 1.684990 | 7.9e-05 |
| SRIME | TDM | 1.786145 | 2.505483 | 3.671453 | 0.00041 |
| MRIME | | **1.235854** | 1.240255 | 1.276928 | 1.26e-05 |
| DO | | 1.282257 | 1.947251 | 2.279401 | 0.00021 |
| NGO | | 1.659848 | 1.954069 | 2.203206 | 0.00011 |
| CLRao-1 | | 1.237706 | 1.365158 | 1.703721 | 9.7e-05 |

Then, Fig. 8 displays the RMSE box plots of the three best algorithms for the SDM, DDM and TDM parameter extraction. Compared to MRIME and CLRao-1, TEMRIME is more robust with fewer outliers across all the PV models.

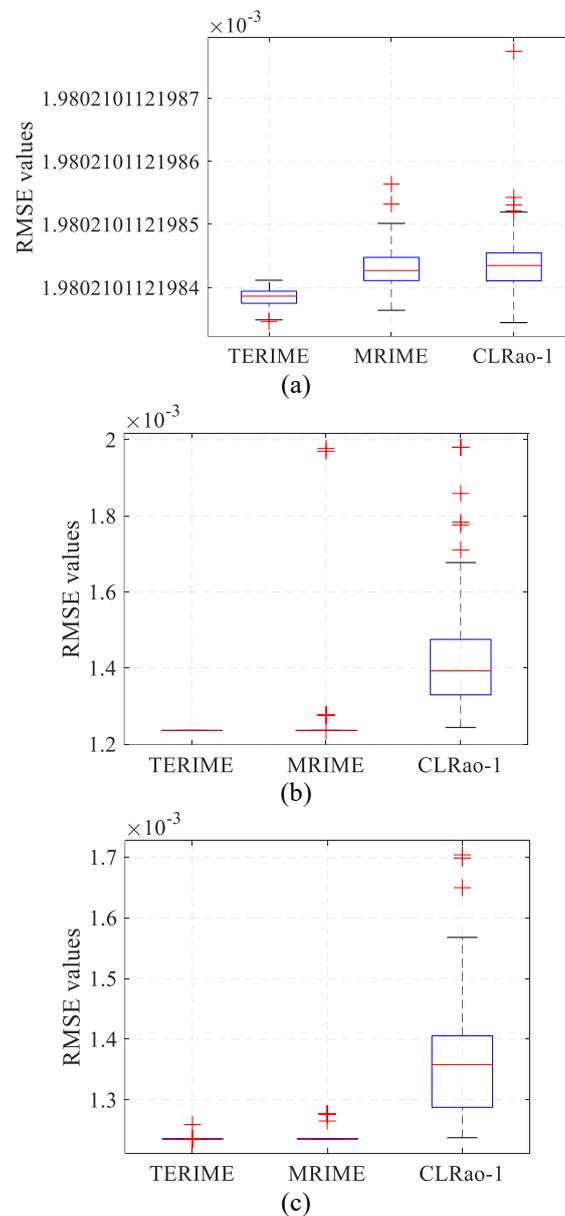

Fig. 8 RMSE box plots for three excellent algorithms for parameter extraction of PWP 201: (a) SDM; (b) DDM; (c) TDM.

Besides, the average convergence of all the algorithms for the SDM, DDM and TDM parameter extraction is illustrated in Fig. 9. As seen in Fig. 9, the following conclusions can be drawn:

- For the SDM, TERIME has the fastest convergence speed to the global optimum, followed by CLRao-1.
- For the DDM, MRIME converges fastest within the first 20000 iterations. Nevertheless, TERIME outperforms all other algorithms after approximately 40000 iterations.
- For the TDM, MRIME has the fastest convergence before 80000 iterations, followed by CLRao-1. However, TERIME overtakes MRIME after 80000 iterations.
- RIME and SRIME are easily trapped in the local optimum, while DO, NGO and SLCRIME converge slowly and fail to approach the global optima for the DDM and SDM.

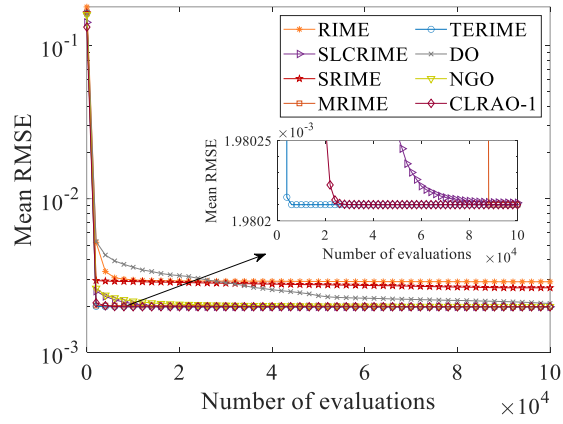

(a)

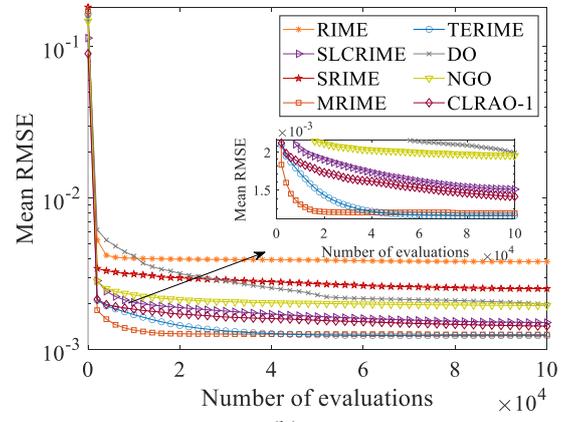

(b)

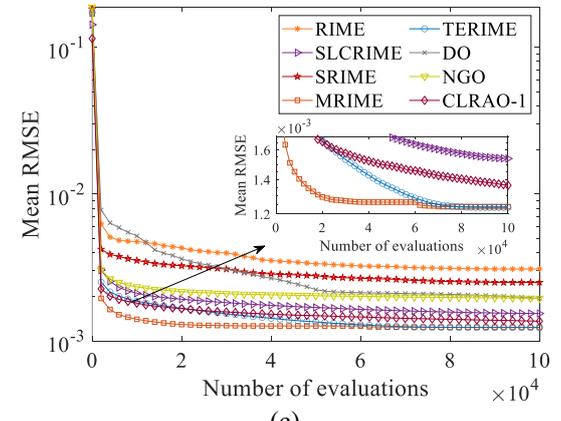

(c)

Fig. 9 Average convergence performance from 100 runs for parameter extraction of PWP 201: (a) SDM; (b) DDM; (c) TDM.

Based on the above results, Fig. 10 illustrates the ranking of all the algorithms for DDM and TDM parameter extraction of PWP 201. As shown in Fig. 10, TERIME is always the best for the parameter identification of PWP 201 in terms of Min, Mean and Max, demonstrating its exceptional performance and robustness.

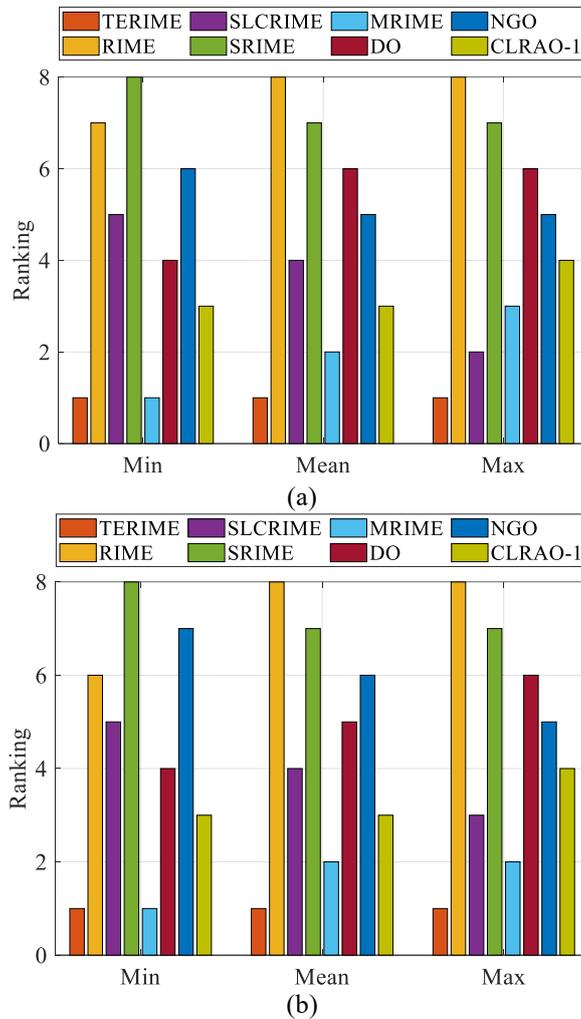

Fig. 10 The ranking of all the algorithms for parameter extraction on PWP 201: (a) DDM; (b) TDM.

Next, we examine the goodness of fit between the calculated data and the measured data. Fig. 11 illustrates the IAEs of SDM, DDM, and TDM for the PWP 201 using parameters obtained by TERIME. It is evident that all the measured data are below 4e-3, irrespective of the PV model type, demonstrating the effectiveness of TERIME. Interestingly, while the IAE for DDM is smaller than that of SDM, it is almost identical to the TDM, indicating that adding more diodes does not always enhance model accuracy.

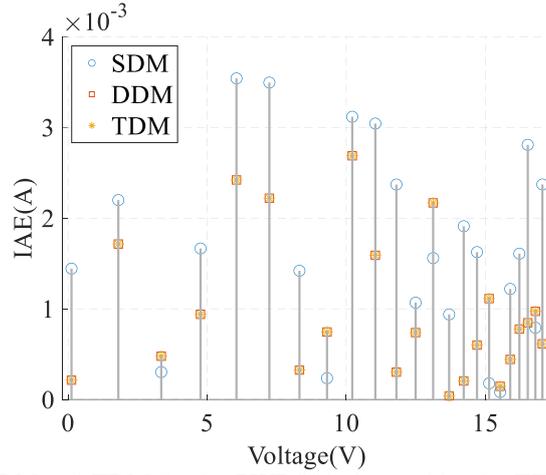

Fig. 11 IAEs of SDM, DDM and TDM for the PWP 201 obtained by the TERIME.

Additionally, the calculated I-V curves for SDM, DDM, and TDM with parameters extracted by TERIME are compared with the measured one in Fig. 12. This comparison indicates that the optimal parameters estimated by TERIME are very close to the actual cell parameters, as the calculated I-V curves align closely with the measurements.

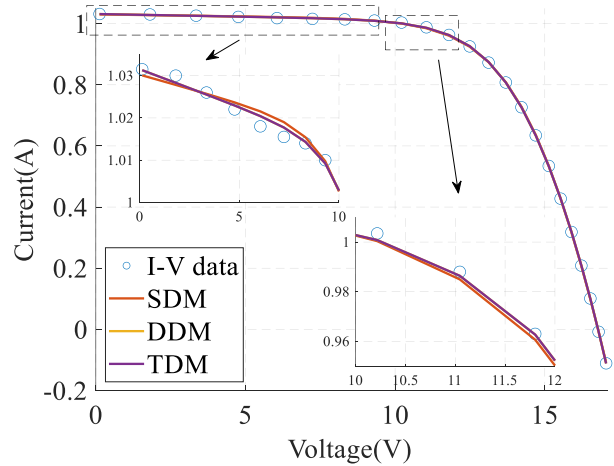

Fig. 12 Comparison of measured I-V curve and calculated ones by the SDM, DDM, and TDM from TERIME for the PWP 201.

5.4 Results of S75 PV Module under Varying Irradiance and Temperature

Based on the above results, the difference in the performance of the meta-heuristic algorithms is mainly in the parameter extraction of DDM and TDM. Therefore, in this subsection, we compare the RMSE results for parameter extraction on DDM and TDM under varying irradiance and temperature. Besides, instead of comparing all the algorithms as in Sections 5.2 and 5.3, we choose MRIME, and CLRao-1 for comparison because they always achieve remarkable performance in the above PV parameter extraction task.

Firstly, the RMSE results for DDM parameter extraction of the S75 under various irradiance and temperature are listed in Table 4, and the corresponding RMSE box plots are shown in Fig. 13. From Table 4, although all the algorithms can give a global optima in 100 runs under all the conditions except

CLRao-1 at 200 W/m² and 25 °C, TERIME has the best Mean, Max and SD under all the conditions, which showcases its superior robustness. Besides, as shown in Fig. 13, TERIME has fewer outliers compared to other algorithms under all the conditions.

Table 4 Comparison of RMSE results from 100 runs for DDM parameter extraction of S75 under varying irradiance and temperature.

| Algorithms | G/ W·m$^{-2}$ | T/ °C | Min / 10$^{-2}$ | Mean / 10$^{-2}$ | Max / 10$^{-2}$ | SD |
|---|---|---|---|---|---|---|
| TERIME |  |  | **0.432351** | **0.432351** | **0.432351** | **1.27e-17** |
| MRIME | 200 | 25 | **0.432351** | 0.432811 | 0.445754 | 1.84e-05 |
| CLRao-1 |  |  | 0.432444 | 0.432621 | 0.432773 | 1.30e-06 |
| TERIME |  |  | **1.102118** | **1.102118** | **1.102118** | **5.58e-17** |
| MRIME | 400 | 25 | **1.102118** | 1.102129 | 1.103145 | 1.03e-06 |
| CLRao-1 |  |  | **1.102118** | 1.102118 | 1.102118 | 6.10e-17 |
| TERIME |  |  | **1.418367** | **1.418367** | **1.418367** | **7.50e-17** |
| MRIME | 600 | 25 | **1.418367** | 1.418368 | 1.418394 | 2.66e-08 |
| CLRao-1 |  |  | **1.418367** | 1.418367 | 1.418367 | 7.96e-17 |
| TERIME |  |  | **1.968935** | **1.968935** | **1.968935** | **9.31e-17** |
| MRIME | 800 | 25 | **1.968935** | 1.968994 | 1.974749 | 5.81e-06 |
| CLRao-1 |  |  | **1.968935** | 1.968935 | 1.968935 | 1.05e-16 |
| TERIME |  |  | 1.962142 | **1.962257** | **1.969512** | **8.14e-06** |
| MRIME | 1000 | 25 | **1.962142** | 1.987169 | 3.430604 | 0.00173 |
| CLRao-1 |  |  | **1.962142** | 2.007249 | 2.088595 | 0.00019 |
| TERIME |  |  | **1.811789** | **1.811789** | **1.811789** | **1.46e-16** |
| MRIME | 1000 | 20 | **1.811789** | 1.839210 | 2.974532 | 0.00129 |
| CLRao-1 |  |  | **1.811789** | 1.859859 | 2.026441 | 0.00066 |
| TERIME |  |  | **1.281773** | **1.281773** | **1.281773** | **1.53e-16** |
| MRIME | 1000 | 40 | **1.281773** | 1.285465 | 1.425530 | 0.00021 |
| CLRao-1 |  |  | 1.281774 | 1.317110 | 1.492325 | 0.00052 |
| TERIME |  |  | **2.578601** | **2.578601** | **2.578601** | **1.24e-16** |
| MRIME | 1000 | 60 | **2.578601** | 2.580017 | 2.682043 | 0.00011 |
| CLRao-1 |  |  | **2.578601** | 2.582320 | 2.881917 | 0.00031 |

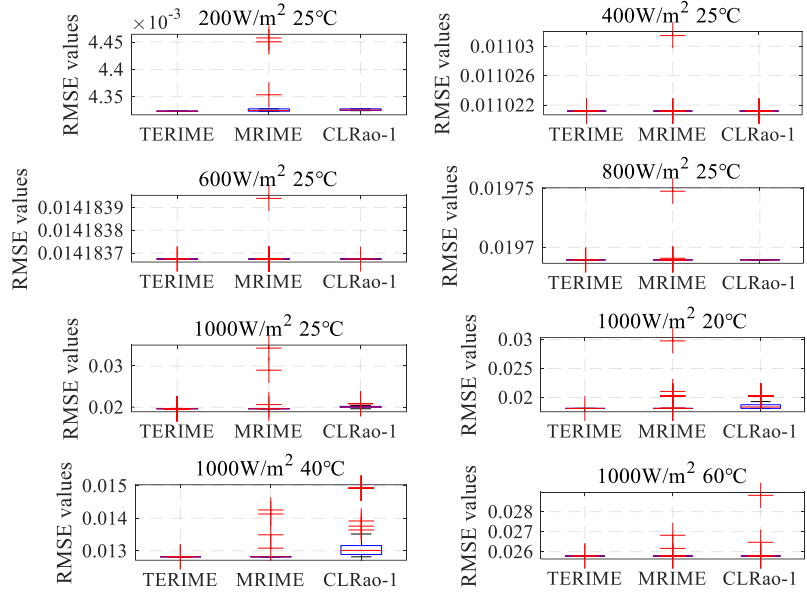

Fig. 13 RMSE box plots for three excellent algorithms for DDM parameter extraction of S75 under varying irradiance and temperature.

Then, the RMSE results for TDM parameter extraction of the S75 under various irradiance and temperature are listed in Table 5, and the corresponding RMSE box plots are shown in Fig. 14. From Table 5, MRIME and TERIME can always give the global optima in 100 runs under all the conditions. Notably, TERIME has the best Max and SD under all the conditions, showing its extraordinary robustness. Besides, as seen in Fig. 14, TERIME has fewer outliers compared to other algorithms under all the conditions.

Table 5 Comparison of RMSE results from 100 runs for TDM parameter extraction of S75 under varying irradiance and temperature.

| Algorithms | G/ W·m$^{-2}$ | T/ °C | Min / 10$^{-2}$ | Mean / 10$^{-2}$ | Max / 10$^{-2}$ | SD |
|---|---|---|---|---|---|---|
| TERIME |  |  | **0.431861** | **0.431948** | **0.431969** | **2.2e-07** |
| MRIME | 200 | 25 | **0.431861** | 0.432303 | 0.432773 | 3.2e-06 |
| CLRao-1 |  |  | 0.432247 | 0.432535 | 0.432773 | 9.3e-07 |
| TERIME |  |  | **1.099577** | **1.099577** | **1.099577** | **6.6e-17** |
| MRIME | 400 | 25 | **1.099577** | **1.099577** | **1.099577** | 2.8e-16 |
| CLRao-1 |  |  | **1.099577** | **1.099577** | **1.099577** | 7.1e-17 |
| TERIME |  |  | **1.418367** | **1.418367** | **1.418367** | **7.8e-17** |
| MRIME | 600 | 25 | **1.418367** | 1.418484 | 1.430003 | 1.1-05 |
| CLRao-1 |  |  | **1.418367** | **1.418367** | **1.418367** | 1.2e-16 |
| TERIME |  |  | **1.968935** | **1.968935** | **1.968935** | **8.7e-17** |
| MRIME | 800 | 25 | **1.968935** | 1.968993 | 1.974749 | 5.8e-06 |
| CLRao-1 |  |  | **1.968935** | 1.998208 | 1.974749 | 5.8e-06 |
| TERIME | 1000 | 25 | 1.950811 | 1.959215 | **1.974729** | **4.0e-05** |
| MRIME |  |  | **1.944691** | **1.957261** | 2.094792 | 0.00015 |
| CLRao-1 | | | 1.957345 | 2.005946 | 2.089124 | 0.00018 |
| TERIME | | | **1.760594** | 1.771689 | **1.811789** | **0.00012** |
| MRIME | 1000 | 20 | **1.760594** | **1.763839** | 1.878742 | 0.00015 |
| CLRao-1 | | | **1.760594** | 1.864216 | 2.975370 | 0.00122 |
| TERIME | | | **1.264338** | **1.268166** | **1.283026** | **3.4e-05** |
| MRIME | 1000 | 40 | **1.264338** | 1.285377 | 1.663468 | 0.00079 |
| CLRao-1 | | | 1.269061 | 1.310599 | 1.663485 | 0.00042 |
| TERIME | | | **2.222116** | **2.222116** | **2.222116** | **1.5e-15** |
| MRIME | 1000 | 60 | **2.222116** | 2.224396 | 2.450037 | 0.00023 |
| CLRao-1 | | | **2.222116** | 2.262531 | 2.579461 | 0.00108 |

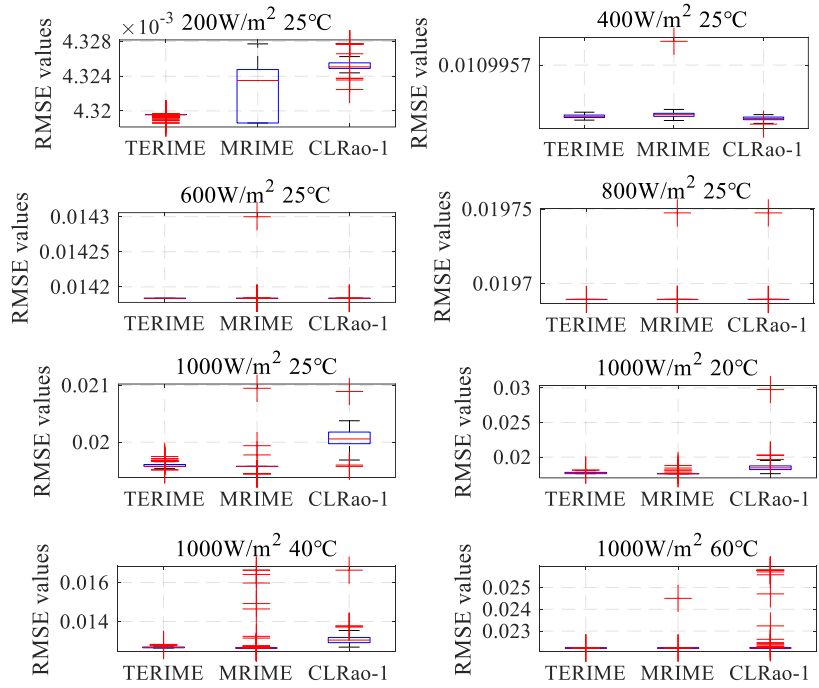

Fig. 14 RMSE box plots for three excellent algorithms for TDM parameter extraction of S75 under varying irradiance and temperature.

Besides, the average rankings of these three algorithms for parameter extraction on S75 across all the environmental conditions are illustrated in Fig. 15. It is shown that TERIME presents the best ranking of Mean and Max for both the DDM and TDM, although the ranking of Min for TDM is slightly inferior to MRIME. This demonstrates that TERIME can give a superior robust solution under varying environmental conditions.

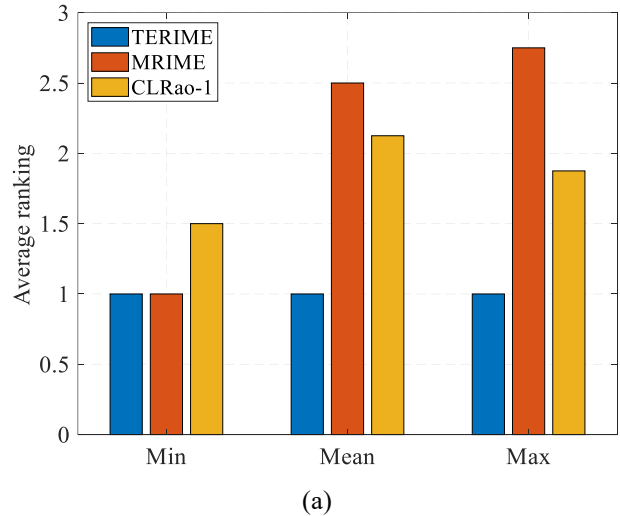

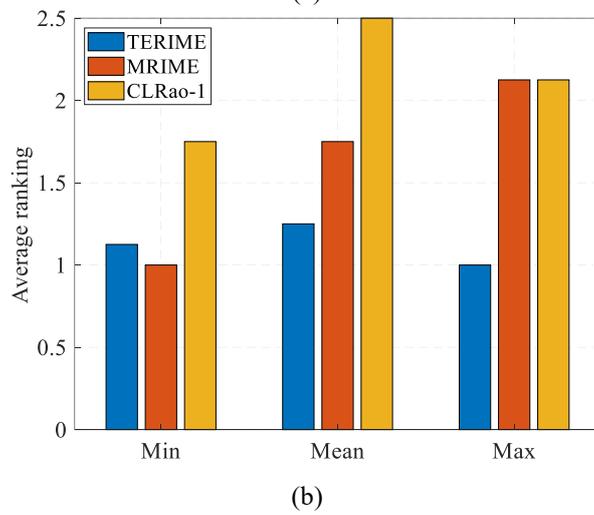

Fig. 15 Average ranking of three excellent algorithms of parameter extraction for S75: (a) DDM; (b) TDM

5.5 Discussions of Parameter Settings

The population size and the number of evaluations significantly influence the performance of meta-heuristic algorithms. Therefore, in this section, we focus on analyzing their effects on the experimental results using the parameter extraction of RTC France as a case study. Again, only the results of DDM and TDM are analyzed.

Firstly, the average convergence performances of all the algorithms under different population sizes on DDM are shown in Fig. 16. For DDM parameter extraction, the following conclusion can be drawn:

- As the population size increases, the convergence speed of TERIME initially accelerates and then decreases, with the optimal speed observed at $N=20$.
- Across all population sizes, TERIME demonstrates superior performance compared to other algorithms at 100000 evaluations of the objective function.
- With a sufficient population size but limited evaluations of the objective function, TERIME performs worse than MRIME. However, the performance of MRIME drops significantly with a small population size (e.g., $N=10$), while TERIME maintains excellent performance.

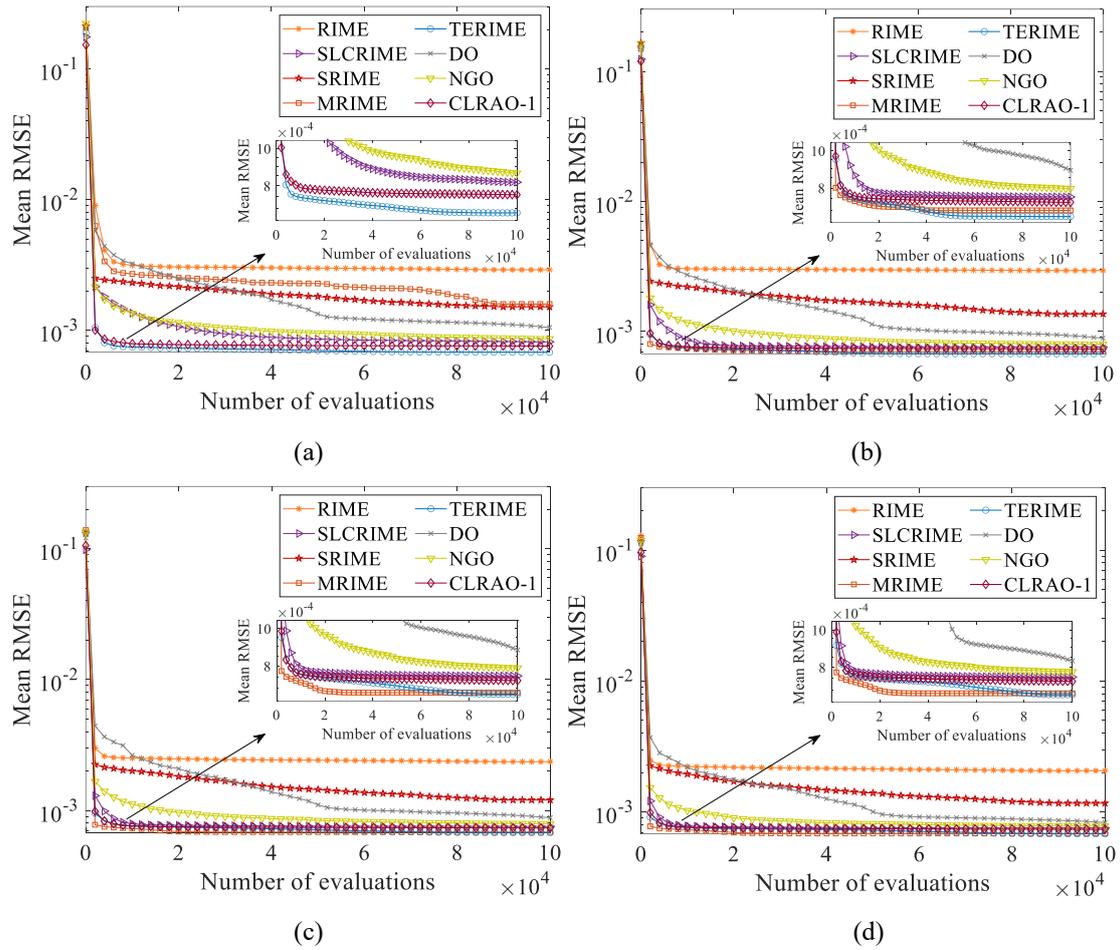

Fig. 16 Average convergence performance from 100 runs for parameter extraction of RTC France on DDM: (a) $N$=10; (b) $N$=20; (c) $N$=30; (d) $N$=40.

Then, the minimum and mean values of RMSE from 100 runs using TERIME for parameter extraction of RTC France on DDM under different population sizes and number of objective function evaluations are illustrated in Fig. 17. It can be seen that when $N$=20, TERIME presents better results compared to other population sizes. As the number of evaluations increases, the performance of TERIME improves. When $N$=20 and $E_{max}$=100000, TERIME gives the best results.

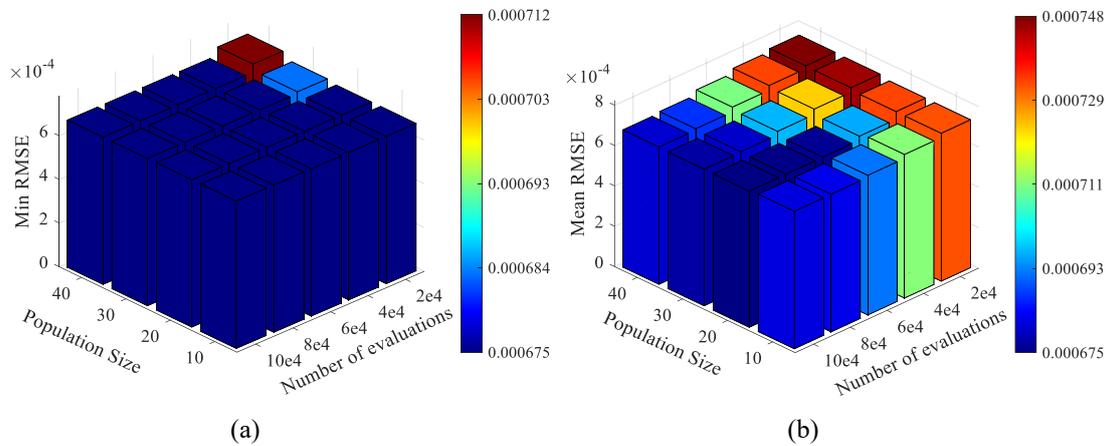

Fig. 17 Minimum and mean values of RMSE from 100 runs using TERIME for parameter extraction of RTC France on DDM under different population sizes and number of objective function evaluations: (a) Min; (b) Mean.

Next, the average convergence performances of all the algorithms under different population sizes on TDM are displayed in Fig. 18. For TDM parameter extraction, we can deduce that:
- As the population size increases, TERIME needs more number of evaluations for convergence.
- When $N$=10 and $N$=20, TERIME demonstrates superior performance compared to other algorithms at 100000 evaluations of the objective function, while as $N$=30 and $N$=40, its performance is inferior to MRIME due to limited number of function evaluations.

Subsequently, the minimum and mean values of RMSE from 100 runs using TERIME for parameter extraction of RTC France on TDM under different population sizes and number of objective function evaluations are illustrated in Fig. 19. It can be seen that as the number of evaluations increases, the performance of TERIME improves. When $N$=20, TERIME presents better Min values compared to other population sizes. When $N$=10, TERIME presents better Mean values. If a robust result is preferred in practical applications, $N$=10 and $E_{max}$=100000 should be selected.

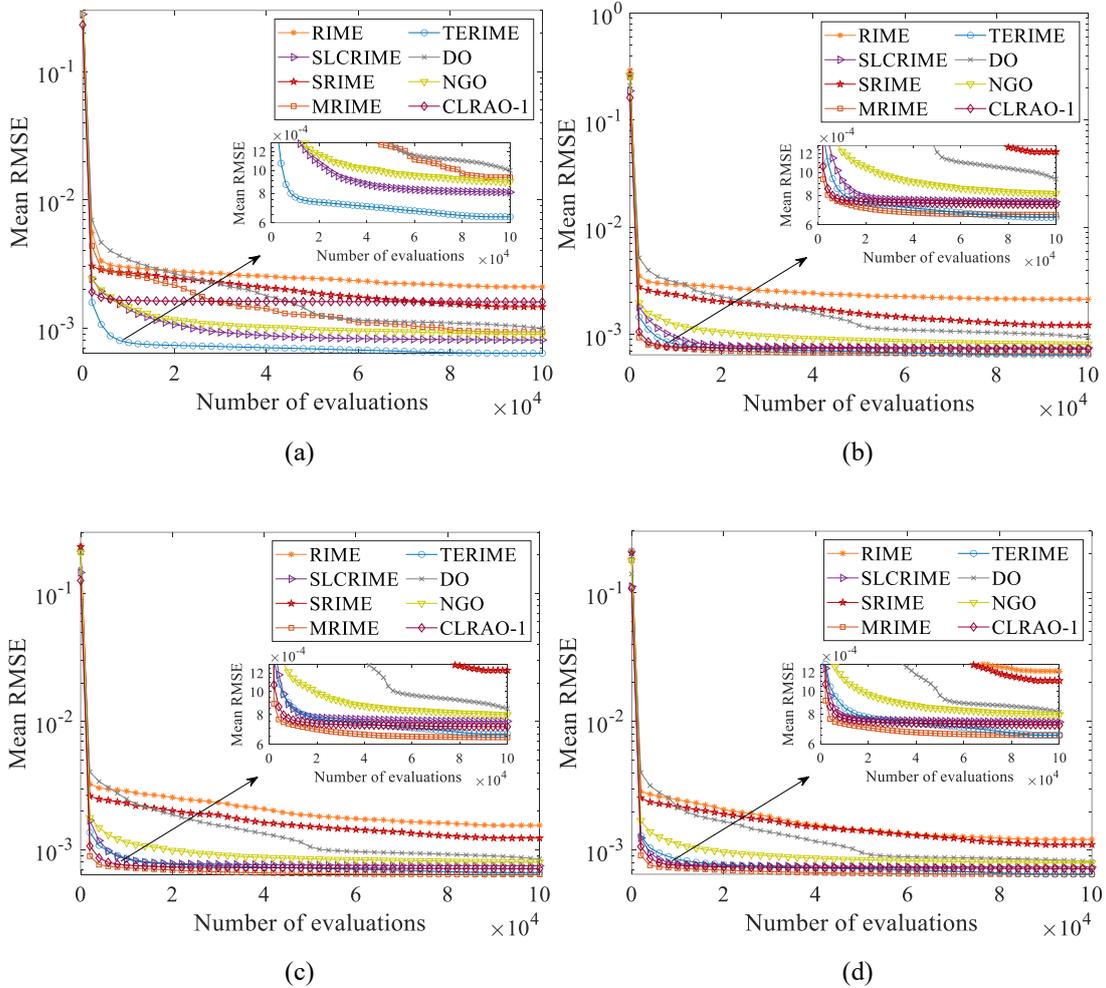

Fig. 18 Average convergence performance from 100 runs for parameter extraction of RTC France on TDM: (a) $N$=10; (b) $N$=20; (c) $N$=30; (d) $N$=40.

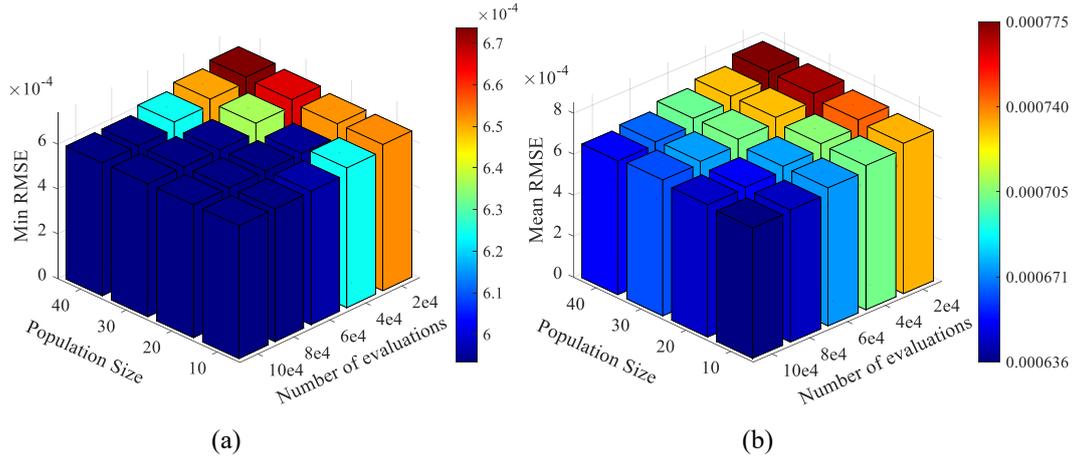

(a)                                          (b)

Fig. 19 Minimum and mean values of RMSE from 100 runs using TERIME for parameter extraction of RTC France on TDM under different population sizes and number of objective function evaluations: (a) Min; (b) Mean.

## 6 Conclusion

In order to tackle the robustness challenge in PV parameter extraction as the complexity of the PV model increases, an improved RIME algorithm called TERIME is proposed. In TERIME, the DE/rand/1 mutation operator is integrated into the exploration phase to enhance population diversity. During the exploitation phase, the crossover strategy and the Gaussian strategy are combined to exchange information among agents randomly and exploit the neighborhood of the current best agent. The results on three PV datasets demonstrate that:

- For datasets 1 and 2, TERIME consistently presents robust solutions as PV model complexity increases, with average RMSE reduction of (3.24%, 2.57%) and (1.70%, 0.34%) over other algorithms for (DDM, TDM) parameter extraction on the RTC France and PWP201, respectively.
- For dataset 3, TERIME provides robust solutions under varying environmental conditions, achieving the best average rankings in Mean and Max values of RMSE for DDM and TDM parameter extraction.
- Based on discussions, when using TERIME for parameter extraction of DDM and TDM, population size is suggested to be set as 10 and 20, respectively.

In summary, TERIME can serve as a reliable optimization tool for accurate parameter identification across various PV models and environmental conditions. Beyond the work of this study, we will extend the proposed exploitation strategy to other meta-heuristic algorithms. Furthermore, dynamic adjustment of the exploitation strategy during iterations is an interesting direction for enhancing the convergence speed of the proposed method in limited evaluations.

## Declarations

Ethics Approval and Consent to Participate

Not applicable

Consent for Publication

All authors have reviewed and approved the final version of this manuscript for publication.

Availability of Data and Material

The datasets and code used during the current study are available at https://github.com/dirge1/TERIME.

Funding

This work was supported by the National Natural Science Foundation of China [grant number 51775020], the Science Challenge Project [grant number. TZ2018007], the National Natural Science Foundation of China [grant number 62073009], the Postdoctoral Fellowship Program of CPSF [grant number GZC20233365], and the Fundamental Research Funds for Central Universities [grant number JKF-20240559].

Conflict of Interest

The authors have no competing interests to declare that are relevant to the content of this article.

Authors' Contributions

**Conceptualization**: Shi-Shun Chen, Yu-Tong Jiang, Wen-Bin Chen, Xiao-Yang Li; **Methodology**: Shi-Shun Chen, Yu-Tong Jiang; **Formal analysis and investigation**: Shi-Shun Chen, Yu-Tong Jiang; **Writing - original draft preparation**: Shi-Shun Chen; **Writing - review and editing**: Shi-Shun Chen, Yu-Tong Jiang, Wen-Bin Chen, Xiao-Yang Li; **Funding acquisition**: Wen-Bin Chen, Xiao-Yang Li; **Resources**: Xiao-Yang Li; **Supervision**: Xiao-Yang Li.

Acknowledgements

We sincerely appreciate the valuable comments and constructive suggestions provided by the anonymous reviewers and the editor, which have greatly contributed to improving the quality of this paper.

Appendix

The short circuit current of the S75 PV module $I_{sc}$ in Table 1 can be calculated by:

$$I_{sc}(G,T) = I_{sc,stc} \cdot \frac{G}{G_{stc}} + k_T \cdot (T - T_{stc}) \tag{33}$$

where $I_{sc,stc}$, $G_{stc}$, and $T_{stc}$ are the short circuit current, irradiance and temperature under standard test conditions (STC), respectively; $G$ represents the operating irradiance; and $k_T$ indicates the temperature coefficient for the short-circuit current. Table 6 displays the parameter values of the S75 in Eq. (33), which are extracted from its datasheet.

Table 6  Parameter values of the S75 extracted from its datasheet.

| Parameters | Values |
| --- | --- |
| $I_{sc,stc}$ / A | 4.7 |
| $k_T$ / mA·°C$^{-1}$ | 2 |
| $G_{stc}$ / W·m$^{-2}$ | 25 |
| $T_{stc}$ / °C | 1000 |